\documentclass[a4paper]{mem}
\usepackage{natbib}
\usepackage{graphicx}
\usepackage[a4paper]{hyperref}
\idline{??}{0}
\begin{document}
   \title{ATLAS and SYNTHE under Linux
}

   \author{L. Sbordone \inst{1,2},
   P. Bonifacio \inst{3},
           F. Castelli \inst{3,4}
          \and
          R. L. Kurucz \inst{5}
}

   \offprints{L. Sbordone}
\mail{ESO - Alonso de Cordova 3107 - Vitacura - Santiago de Chile}


	        \institute{ESO - Santiago de Chile; \email{lsbordon@eso.org}
 \and
   Universita' di Roma 2 ``Tor Vergata'' -  Italy  \and
   INAF - Osservatorio Astronomico di Trieste - Italy  \and
   CNR - Istituto di Astrofisica Spaziale e Fisica Cosmica, Roma - Italy \and
   Harvard - Smithsonian CfA - Cambridge, MA - USA
             }

   \abstract{
   We have successfully ported under GNU Linux ATLAS 9,
the widely used stellar atmosphere modeling code,
as well as both the SYNTHE suite of programs,
its ``companion'' for spectral synthesis, and WIDTH,
used to derive chemical abundances from equivalent widths of
spectral lines. The porting has been realized by using the
Intel Fortran Compiler. Our aim was to port the codes with the
minimum possible amount of modifications:
full compatibility with the VMS version has been maintained,
along with all the codes functionalities.
Dramatic improvement in calculation speed with
respect to the VMS version has been achieved.
The full suite of codes is intended to be freely available to anyone.
   }
   \authorrunning{L. Sbordone et al.}
   \titlerunning{ATLAS and SYNTHE under Linux}
   \maketitle
%

\section{Introduction}

Version 9 of the ATLAS code \citep{kurucz70,kurucz93} is widely used for the production of LTE one-dimensional atmosphere models. Its fields of application extend from the modeling of stellar atmosphere for the purpose of stellar abundance analysis, to the computations of extended grids of colors for the production of synthetic CMDs, to the computation of large grids of synthetic spectra for stellar population synthesis.
ATLAS is complemented by a  suite of  codes for spectrum synthesis (to which we shall refer collectively as SYNTHE) and by WIDTH, a code to derive chemical abundances from the equivalent widths (EW) of observed lines.
Since now, the original versions of the codes were developed only for VMS machines, although some unix portings exist. Most of these portings, nevertheless, depart significantly from the original code, and have tipically been adjusted to the specifical needs of the people involved in the porting.
We undertook the porting under GNU Linux  of ATLAS 9, WIDTH and SYNTHE packages with two main goals in mind:
\begin{itemize}
\item{To make the codes available under Linux, an operating system nowadays far more widespread than VMS, and running on much less expensive machines;}
\item{To port the code introducing the smallest possible alterations in it, leaving unchanged input/output formats and functionalities of the programs. This would allow a smooth integration with the already existing line databases, opacities and model grids. Moreover, this would allow to move effortlessly to Linux any new modification in the VMS code.}
\end{itemize}


\begin{table}
\begin{center}
{\small
\begin{tabular}{lcc}
\hline
\\
                    & {\bf ATLAS } &  {\bf SYNTHE} \\
                    &   s          &  s \\
{\bf VMS}           & 478~(3.54)   & 69                      \\
{\bf Pentium 4}     & 177~(1.31)   & 10                      \\
{\bf Pentium M}     & 122~(0.9)    & 10                      \\
\hline
\end{tabular}
}
\end{center}
\caption{Execution time comparison for ATLAS 9, calculated on 135 iterations of a 72 layers model (time per iteration within parentheses), and for the calculation of a 5 nm synthetic spectrum at resolution 600000 with SYNTHE. Systems are: VMS: AlphaServer 800 5-500; Pentium 4: 1.9 GHz 768 Mb RAM kernel 2.4.18-3 (Red Hat 7.3)  IFC 7.0; Pentium M: 1.6 GHz, 512 Mb RAM kernel 2.4.22-1.2115.nptl (Fedora Core 1) IFC 8.0. All times in seconds. }
\label{times}
\end{table}

\section {Porting with Intel Fortran Compiler}
ATLAS 9 and the other codes in the suite have been developed along some 40 years. They are mostly written in Fortran IV. Thus, a very good set of backward compatibility options were needed in the compiler, in order to minimize the needed changes.
We choose to employ the Intel Fortran Compiler (IFC), a F90-95 (with proprietary extensions) compiler whose Linux version is licensed free of charge for research application. Versions 7.0 and 8.0 have been used.
The changes to the codes were minimal. Among the most important:
\begin{itemize}
\item{Some {\tt FORMAT} statement syntaxes have been updated;}
\item{The syntax of the {\tt OPEN} commands have been modified to fit the new UNIX execution scripts;}
\item{Some variables have been redefined to keep them in double precision throughout the various routines;}
\item{Some big data tables (e.g. partition functions, integration matrices) were originally written in form of huge {\tt DATA} blocks. IFC appears to have a limit to the size of DATA blocks, so we needed either to move them to files or to split them into smaller {\tt DATA} blocks to
avoid compilation problems. This lead to the need for ATLAS to read a further input file ({\tt PFIRON.DAT}) with respect to the VMS version;}
\item{Molecular input data in SYNTHE have been passed to ASCII format (from the VMS binary of the original). The program reading these files has been changed accordingly.}
\end{itemize}

      \begin{figure}
   \centering
   \includegraphics[width=7cm]{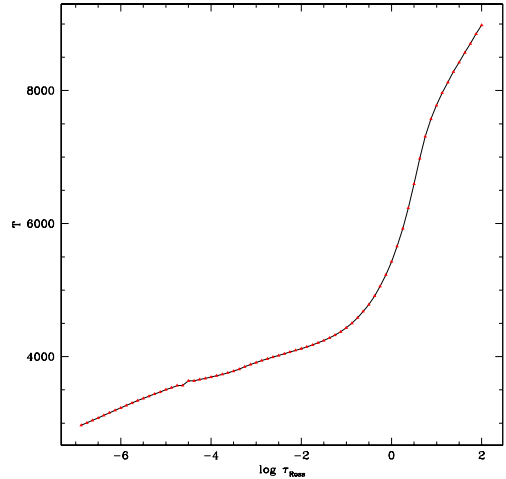}
      \caption{Plot of T against $\log (\tau_{Ross})$ along the atmosphere of a star of $T_{eff}$ = 5000 K, log g = 2.5, $[Fe/H]$=-0.5, as produced by the Linux ported code (solid line) and the original VMS code (triangles). }
         \label{temperature}
   \end{figure}

   \begin{figure*}
   \centering
   \includegraphics[width=13cm]{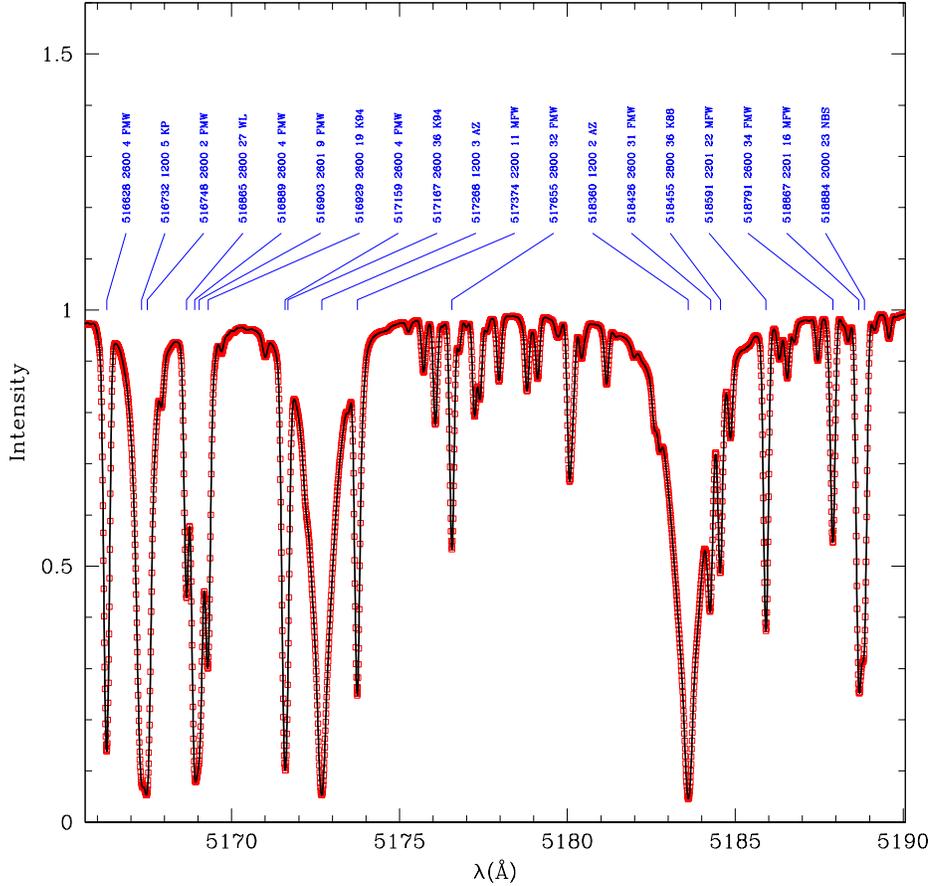}
      \caption{Synthetic spectrum for the above mentioned star around the Mg b triplet. Open squares is VMS original code, solid line Linux code. The resolution of the calculation is 600000, then a gaussian broadening (7 km/s FWHM) has been added to simulate the output of a spectrometer with resolution of about 43000. An example of line label: 518360 = 518.360 nm; 1200 = Mg (12 atomic number) neutral (00); 2 = $2\%$ residual intensity at line center (unbroadened spectrum); AZ is a code for the line $\log(gf)$ source.
              }
         \label{syntheses}
   \end{figure*}

   \begin{figure}
   \centering
   \includegraphics[width=7cm]{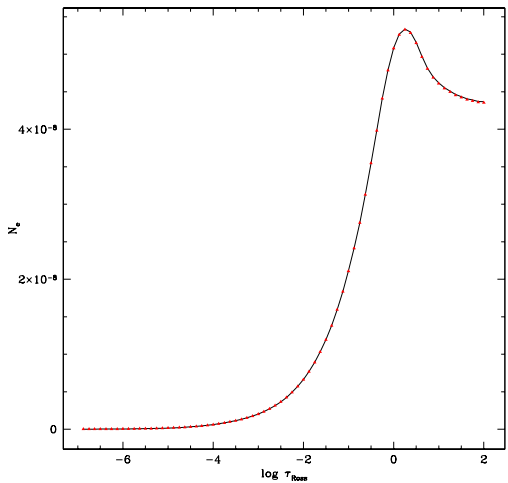}
      \caption{Same as in fig. \ref{temperature}, but now comparing electron number densities.}
         \label{elecdens}
   \end{figure}


%
%

\section{Scripts and input data}

Original VMS execution scripts have obviously been rewritten as C-- shell scripts, but the program sequence and inputs remain unchanged (except for the need to read {\tt PFIRON.DAT}).
All the ASCII inputs (models, line lists, Rosseland opacities) remaining the same, we needed to ``translate'' the Opacity Distribution Functions (ODF), since they were written in VMS binary form, unreadable under Linux. They have been copied in ASCII format and then rewritten in Linux binary (ODFs are bulky and the binary format saves significant disk space).
We have ported program versions for both the ``old'' \citet{kurucz93} ODFs and the ``new'' \citet{castelli03} ODFs.

\section{Performance}
The results obtained with the ported version of the codes are, as can be seen in fig. 1 through 3, indistinguishable from the ones produced by the VMS version.

The porting leads to a dramatic increase in performance with respect to the VMS version. Examples of ATLAS and SYNTHE execution times (WIDTH is a  very fast program, with execution times of about 1 - 2 seconds) are summarized in table \ref{times}. Listed execution times are indicative since they also depend on disk throughtput. As can be seen, both atmosphere modeling and spectral synthesis can now be performed in very short times on a mainstream laptop computer (the system labeled as ``Pentium M'' in table \ref{times} is
actually a laptop). 
This would allow any user to 
calculate directly any atmosphere he needs, 
without having to rely on precalculated grids.

Moreover, together with the use of Linux supercomputer architectures such as Beowulf, this allows to use this programs in the framework of large automated spectral analysis codes \citep[see for instance][]{BC03} conceived to analyze semi-automatically the high amount of data produced by high multiplexity spectrographs such as VLT-FLAMES. The high speed at which both the atmosphere modeling and the spectral synthesis steps can be performed, allows to produce large amounts of Monte Carlo tests to assess the  error budget of the derived abundances.

\section{Availability}
The suite, as the original Kurucz one, is intended to be freely available to everyone interested in it. The full disk space required is about 700 Mb, for a typical set of ODFs of the BIG type (for model computation), line lists and molecular data. For the time being, no webpage has been set to download the codes from, so interested people are encouraged to contact us to obtain the codes and any information related to them (email {\tt lsbordon@eso.org} or {\tt bonifaci@ts.astro.it} ).

Anyone using the codes for scientific publications is only requested to cite this work along with \citet{kurucz93}. Nevertheless, we plan to publish it on the Web  as soon as possible, along with the proper documentation.

\bibliographystyle{aa}

\end{document}